\documentclass{article}
\usepackage{amsmath,amsthm}
\usepackage{amssymb,latexsym}
\usepackage[mathscr]{eucal}
\usepackage{graphics}
\usepackage{array}

\DeclareMathSizes{13}{13}{9}{8}

\newcommand{\pa}{\partial}

\newcommand{\al}{\alpha}
\newcommand{\dpr}{^{\prime\prime}}

\newcommand{\pr}{^\prime}
\newcommand{\rp}{\right)}
\newcommand{\lp}{\left(}
\newcommand{\rb}{\right]}
\newcommand{\lb}{\left[}
\newcommand{\rel}{\right\}}

\newcommand{\beq}{\begin{equation}}
\newcommand{\eq}{\end{equation}}
\newcommand{\bfv}{{\bf v}}

\newcommand{\bfj}{{\bf J}}
\newcommand{\bfb}{{\bf B}}

\newcommand{\bfe}{{\bf E}}
\newcommand{\bfihat}{\hat {\bf i}}
\newcommand{\na}{\nabla}
\newcommand{\ti}{\times}

\newcommand{\bfve}{\bfv_e}
\newcommand{\bfvi}{\bfv_i}
\newcommand{\vet}{\frac{\pa\bfve}{\pa t}}
\newcommand{\vit}{\frac{\pa\bfvi}{\pa t}}
\newcommand{\pe}{p_e}
\newcommand{\pui}{p_i}
\newcommand{\lc}{\frac{1}{c}}

\usepackage{graphicx}
\graphicspath{{converted_graphics/}}
\begin{document}
\title{Parker Problem in Hall Magnetohydrodynamics}  \author {Bhimsen K. Shivamoggi \footnote {\large Permanent Address: University of Central Florida, Orlando, FL 32816-1364}\\Los Alamos National Laboratory\\ Los Alamos, NM 87545\\}     

\date{}          
\maketitle
\date{}          
\maketitle

\large{\bf Abstract}

Parker problem in Hall magnetohydrodynamics (MHD) is considered. Poloidal shear into the toroidal flow generated by the Hall effect is incorporated. This is found to lead to a {\it triple deck} structure for the Parker problem in Hall MHD, with the magnetic field falling off in the intermediate Hall-resistive region more steeply (like \normalfont $1/x^3$) than that (like \normalfont$1/x$) in the outer ideal MHD region.

\pagebreak

\section{Introduction}      

Numerical simulations (Biskamp \cite{Bis}) suggested that magnetic reconnection in resistive MHD can be driven by a magnetic flux pile-up. In this process, magnetic field builds up upstream of a Sweet \cite{Swe}-Parker \cite{Par1} current sheet - this field build-up strenghtens as the resistivity is decreased,  which leads to an increase in the outflow downstream of the current sheet. This provides for a remedy for the Sweet-Parker  {\it``bottleneck"}\footnote{ \large In the Sweet-Parker model, the {\it bottleneck} in the outflow is produced by slowly moving ions under the influence of a very small transverse magnetic field in the elongated diffusion region.} limiting the outflow and enables reconnection to proceed at the externally imposed rate. However, as the resistivity is decreased further, the development of large magnetic pressure gradients upstream of the current sheet opposes the ion inflow (Knoll and Chacon \cite{Kno1}) which is the only means in resistive MHD to transport magnetic flux into the reconnection layer. The magnetic flux transport into the reconnection layer (and hence the reconnection rate) is reduced - the so-called {\it pressure problem} (Clark \cite{Cla}). The Hall effect (Sonnerup \cite{Son}) can overcome the {\it pressure problem} (Dorelli and Birn \cite{DorB}, Knoll and Chacon \cite{Kno2}) thanks to the decoupling of electrons from ions on length scales below the ion skin depth $d_i$. So, if the reconnection-layer width is less than $d_i$, the electron inflow can keep on going which transports the magnetic flux into the reconnection layer and hence reduces the flux pile-up. Dorelli \cite{Dor} considered the role of Hall effect in flux pile-up driven anti-parallel magnetic field merging and gave analytical solutions of the resistive Hall MHD equations describing stagnation-point flows in a thin current sheet - Parker Problem (Parker \cite{Par2}). However Dorelli's solution for the Hall regime turned out to be basically Parker's solution for the ideal MHD regime - indeed, the Hall effect can be transformed away from Dorelli's solution by suitably redefining the velocity gradient associated with the stagnation-point flow. A more complete formulation fo the Parker problem in Hall MHD is therefore in order - this is the objective of this paper. 

\section{Governing Equations for Hall MHD}
Consider an incompressible, two-fluid, quasi-neutral plasma. The governing equations for this plasma dynamics are (in usual notation) - 

\beq
nm_e \lb \vet + (\bfve\cdot\na)\bfve\rb = -\na\pe-ne (\bfe+\lc\bfve\ti\bfb) + ne\eta \bfj
\eq

\beq
nm_i \lb \vit + (\bfvi\cdot\na)\bfvi\rb = -\na\pui+ne(\bfe+\lc\bfvi\ti\bfb)-ne\eta\bfj
\eq

\begin{align}
\na\cdot\bfve &=0\\
\na\cdot\bfvi &=0\\
\na\cdot\bfb &=0\\
\na\ti\bfb &=\lc\bfj\\
\na\ti\bfe &=-\lc\frac{\pa\bfb}{\pa t}
\end{align}

where,
\beq
\bfj\equiv ne(\bfvi - \bfve).
\eq

Neglecting electron inertia ($m_e \rightarrow 0$), equations (1) and (2) can be combined to give an ion equation of motion - 

\beq
nm_i \lb \vit +(\bfvi\cdot\na)\bfvi\rb = -\na(p_i + p_e) +\lc \bfj\ti\bfb
\eq

\noindent
and a generalized Ohm's law - 

\beq
\bfe + \lc \bfv_i\ti\bfb = \eta\bfj +\frac{1}{nec}\bfj\ti\bfb.
\eq

Non-dimensionalize distance with respect to a typical length scale $a$, magnetic field with respect to a typical magnetic field strength $B_0$, time with respect to the reference Alfv\'en time $\tau_A \equiv a/V_{A_0}$ where $V_{A_0} \equiv B_0/\sqrt{\rho}$ and $\rho \equiv m_i n$, and introduce the magnetic and velocity stream functions according to

\beq
\left.
\begin{array}{l}
\bfb = \na\psi \ti\bfihat_z + b\bfihat_z\\
\bfvi = \na\phi \ti\bfihat_z + w\bfihat_z
\end{array}\rel
\eq

\noindent
and assume the physical quantities of interest have no variation along the $z$-direction. Equations (9) and (10), then yeild

\beq
\frac{\pa\psi}{\pa t} + [\psi,\phi] + \sigma [b,\psi] = \eta \na^2\psi
\eq

\beq
\frac{\pa b}{\pa t} + [b,\phi] + \sigma [\psi,\na^2\psi] +[\psi, w] = \eta\na^2 b
\eq

\beq
\frac{\pa w}{\pa t} + [w,\phi] = [b,\psi]
\eq

\noindent
where,

\beq
[A,B] \equiv \na A \ti \na B\cdot\bfihat_z,\ 
\sigma \equiv \frac{d_i}{a},\ \hat\eta \equiv \eta c^2 \tau_A / a^2.\notag
\eq

\section{Parker Problem in Hall MHD}

Consider a stagnation-point flow at a current sheet separating plasmas of opposite magnetizations (Parker \cite{Par2}) in Hall MHD, governed by equations (12) - (14). Let us assume that the magnetic field lines are straight and parallel to the current sheet, with the current flowing in the z-direction. Here, pure resistive annihilation without reconnection of anti-parallel magnetic fields (in the x,y-plane) occurs. Specifically, consider a unidirectional magnetic field 

\beq
\bfb=B(x)\bfihat_y
\eq

\noindent
with the boundary condition - 

\beq
B(0) = 0
\eq

\noindent
which is carried toward a neutral sheet at $x=0$ by a stagnation-point flow  

\beq
\bfvi = -ax\bfihat_x + ay \bfihat_y + w\bfihat_z.
\eq

Noting that the process in question is steady and that the magnetic field is prescribed as in (15), equations (12) - (14) become

\beq
E + \frac{\pa\psi}{\pa x}\frac{\pa\phi}{\pa y} - \sigma \frac{\pa b}{\pa y}\frac{\pa\psi}{\pa x} = \hat\eta \frac{\pa^2\psi}{\pa x^2}
\eq

\beq
\frac{\pa b}{\pa x}\frac{\pa\phi}{\pa y} - \frac{\pa b}{\pa y}\frac{\pa \phi}{\pa x} + \frac{\pa \psi}{\pa x}\frac{\pa w}{\pa y} = \hat\eta\na^2 b
\eq

\beq
\frac{\pa w}{\pa x} \frac{\pa \phi}{\pa y} - \frac{\pa w}{\pa y}\frac{\pa \phi}{\pa x} + \frac{\pa \psi}{\pa x}\frac{\pa b}{\pa y} = 0
\eq

\noindent
where,

\beq
E\equiv\frac{\pa\psi}{\pa t}\notag
\eq

Dorelli \cite{Dor} looked for a solution of equations (18) - (20) of the form 

\begin{subequations}
\beq
b=yf(x),
\eq
\beq
w=\al g_1(x)
\eq
\end{subequations}

\noindent
where $\al$ is a constant. But, this solution restricts the role played by the Hall effect in the process in question - indeed, the Hall effect can be transformed away from Dorelli's solution by suitably redefining the velocity gradient $a$ associated with the stagnation-point flow. One way to remedy this situation is to incorporate a poloidal shear into the toroidal flow according to 

\beq
w=\al g_1(x) + \frac{\beta}{2} y^2g_2 (x)
\eq

\noindent
where $\beta$ is a constant characterizing this poloidal shear.
Using (21a) and (22), equation (20) gives

\beq
(\al g\pr_1 + \frac{\beta}{2} y^2g\pr_2) (-ax) - (\beta y g_2)(-ay) + \frac{\pa \psi}{\pa x}f\pr = 0
\eq

\noindent
from which, we obtain

\beq
(-ax)(\al g\pr_1) = -\frac{\pa \psi}{\pa x}f\pr
\eq

\beq
g_2(x) = x^2.
\eq

Using (21a),(22),(24) and (25), equation (19) then gives 

\beq
y f\pr\cdot(-ax)-f\cdot(-ay)+\frac{\pa\psi}{\pa x}\cdot(\beta y x^2) = \hat\eta y f\dpr
\eq

\noindent
or

\beq
f\dpr + \frac{a}{\hat\eta} (x f\pr - f) = \frac{\beta}{\hat \eta} x^2 \frac{\pa \psi}{\pa x}.
\eq

Recognizing that the Hall effect becomes important away from the current sheet at $x=0$ (in what was called the {\it ``intermediate"} layer by Terasawa \cite{Ter} in his investigation of the Hall resistive tearing mode), a reasonable approximate solution of equation (27) is 

\beq
f(x) \approx Ax - \frac{\beta}{a} x^2 \frac{\pa \psi}{\pa x}
\eq

\noindent
where $A$ is a constant.

Using (21a) and (28), equation (18) gives

\beq
-E + \frac{\pa \psi}{\pa x} \cdot (-ax) - \sigma \lp Ax - \frac{\beta}{a} x^2 \frac{\pa \psi}{\pa x}\rp \frac{\pa \psi}{\pa x} = \hat\eta \frac{\pa^2\psi}{\pa x^2}\notag
\eq

\noindent
or

\beq
E=\hat\eta B\pr + (a + \sigma A) xB + \sigma \frac{\beta}{a} x^2B^2 
\eq

\noindent
which is a generalization of Parker's equation to Hall MHD.

It is important to note that equation (29) shows that the {\it triple-deck} structure (borrowing the terminology from boundary layer theory in fluid dynamics) in Hall resistive HMD (originally pointed out by Terasawa \cite{Ter}) is operational for the Parker problem in Hall MHD, as to be expected. Thus, we have 

\begin{itemize}
  \item[(i)] a resistive region near the current sheet at $x=0$,
  \item[(ii)] an ideal MHD region away from the current sheet at $x=0$,
  \item[(iii)] a Hall resistive region in between (i) and (ii) - called the "intermediate" region by Terasawa \cite{Ter}.
\end{itemize}  

In the resistive region, equation may be approximated by 

\beq
E\approx \hat\eta B\pr
\eq

\noindent
which gives Parker's solution - 

\beq
B \approx \frac{E}{\hat\eta}x.
\eq

In the ideal MHD region, equation (29) may be approximated by

\beq
E\approx (a + \sigma A) xB+\sigma \frac{\beta}{a}x^2B^2
\eq

which gives the modified Dorelli's solution - 

\beq
B\approx \lb\frac{\sigma\beta}{2a(a+\sigma A)}\rb\lb-1+\sqrt{1+\frac{4E(a+\sigma A) a^2}{\sigma^2\beta^2}}\rb\frac{1}{x}.
\eq

In the Hall resistive region, equation (29) may be approximated by

\beq
E\approx \hat\eta B\pr + \sigma\frac{\beta}{a}x^2B^2
\eq
 
\noindent
which gives

\beq
B\approx\lp\frac{3\hat\eta a}{\sigma\beta}\rp\frac{1}{x^3}.
\eq

The triple-deck structure given by (31), (33) and (35) is shown in Figure 1. 

\indent Noting, from (21a) and (28), that

\beq
b = y(Ax + \frac{\beta} {a} x^2B)
\eq
\noindent
one observes that both in the 
\begin{itemize}
\item  Hall resistive region: $B\sim1/x^3$,
\item  Ideal MHD region: $B\sim1/x$,
\end{itemize}
b has a \textit{quadrupolar} structure. So, Hall effects materialize only via their signature - the \textit{quadrupolar} out-of-plane magnetic field pattern.

\section{Discussion}

In recognition of the fact that a signature of the Hall effect is the generation of out-of-plane ``separator" components of the magnetic and velocity fields, a more accurate representation of the latter appears to be in order for a more complete formulation of the Parker problem. In this paper, this is accomplished by incorporating poloidal shear into the toroidal flow. This is found to lead to a {\it triple-deck} structure for the Parker problem in Hall MHD, in accordance with the idea originally put forward by Terasawa \cite{Ter}. The magnetic field is found to fall off in the intermediate Hall-resistive region more steeply (like $1/x^3$) than that (like $1/x$) in the outer ideal MHD region.

\section{Acknowledgments}

I acknowledge with gratitude the stimulating interactions with Dr. Luis Chacon that led to this work. My thanks are due to Dr. Mike Johnson for stimulating discussions.

\hspace{-.1in}
\begin{center}
  \scalebox{0.7}{\includegraphics{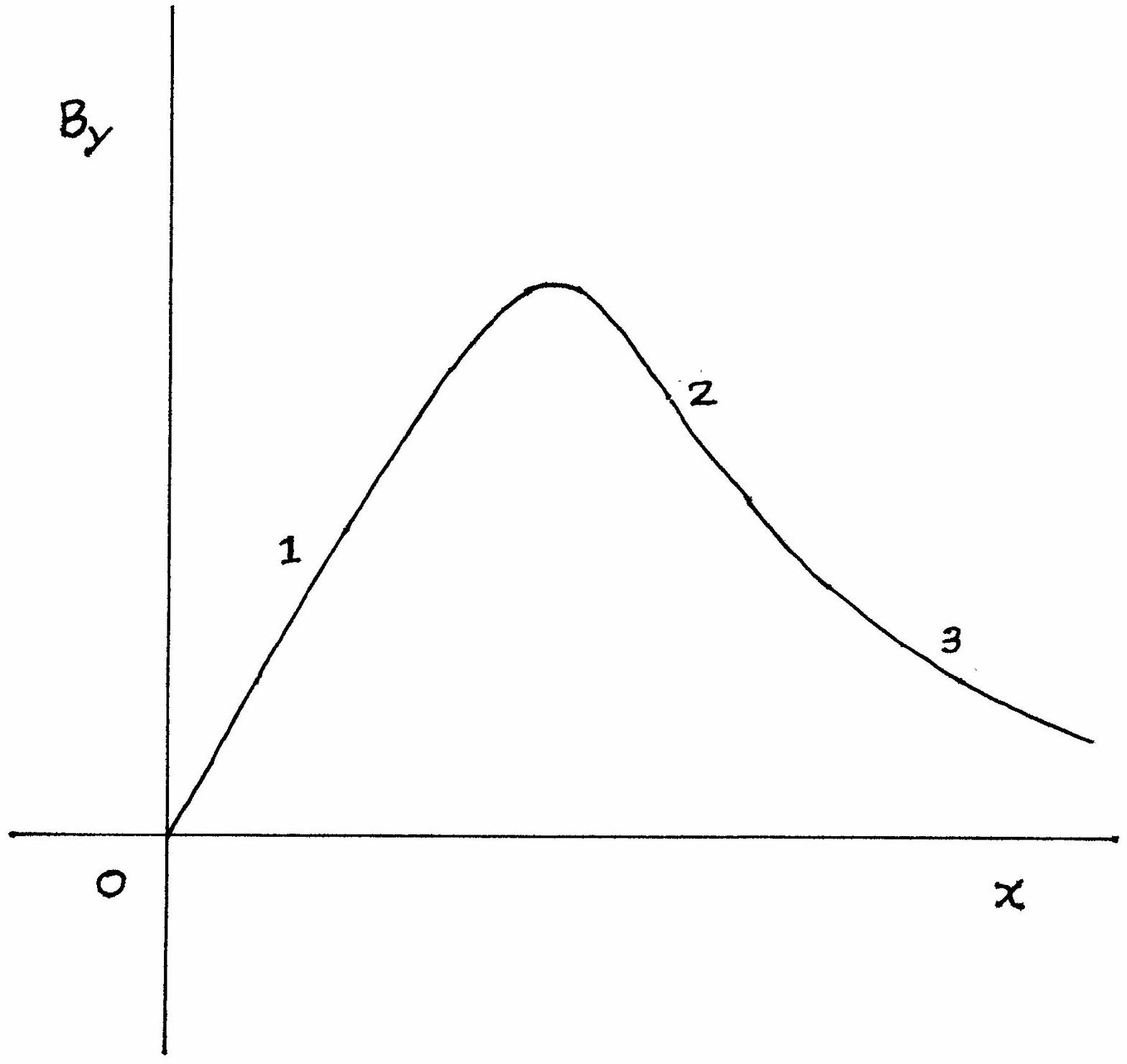}}
\end{center}

Figure 1. Magnetic field profile for Parker Problem in Hall MHD.

\begin{equation}1:\cdot  B\sim x \notag
\end{equation} 

\begin{equation}2: \cdot  B\sim \frac{1}{x^3}\notag
\end{equation}

\begin{equation}3:\cdot   B\sim\frac{1}{x}\notag
\end{equation}
\end{document}